\documentclass{icrc2009}

\usepackage{graphicx}                  
\usepackage[caption=false]{caption}    
\usepackage[font=footnotesize]{subfig} 
\usepackage{fixltx2e}
\usepackage{url}

\newcommand{\shorttitle}[1]%
{\markboth{Proceedings of the 31\MakeLowercase{$^{st}$} ICRC, {\L}\'{o}d\'{z} 2009}{#1} }

\newcommand{\gammaray}{$\gamma$-ray~}
\newcommand{\gammarays}{$\gamma$-rays~}
\newcommand{\hess}{H.E.S.S.~}

\newcommand\apj{{\emph{ApJ}}}
\newcommand\aj{{\emph{AJ}}}
\newcommand\aap{{\emph{A\&A}}}
\newcommand\nat{{\emph{Nature}}}

\begin{document}
\title{Extending the H.E.S.S. Galactic Plane Survey}

\author{\IEEEauthorblockN{Ryan C.G. Chaves\IEEEauthorrefmark{1}
                          on behalf of the H.E.S.S. Collaboration}
                          \\
\IEEEauthorblockA{\IEEEauthorrefmark{1}Max-Planck-Institut f\"ur Kernphysik, P.O. Box 103980, D 69029
Heidelberg, Germany}}


\shorttitle{R.C.G. Chaves. H.E.S.S. Galactic Plane Survey}
\maketitle

\begin{abstract}
The High Energy Stereoscopic System (H.E.S.S.), located in the Khomas Highlands of Namibia, is an array of four imaging
atmospheric-Cherenkov telescopes designed to detect \gammarays in the very-high-energy (VHE; E~$>$~100~GeV) domain.  It is
an ideal instrument for surveying the Galactic plane in search of new sources of VHE $\gamma$-rays, due to its location in
the Southern Hemisphere, its excellent sensitivity, and its large field-of-view.
The \hess Galactic Plane Survey (GPS) began in 2004 
and initial observations of the Galactic plane resulted in the discovery of numerous VHE
\gammaray emitters.  This original \hess GPS covered the inner Galaxy within  
$\pm$~$30$$^{\circ}$~in longitude and $\pm$~$3$$^{\circ}$~in latitude, with respect to the Galactic center.  
In the last four years, the longitudinal extent of the survey has more than doubled, now including the region
$l$~$=$~$275$$^{\circ}$~$-$~$60$$^{\circ}$, and the exposure in the inner Galaxy has also been significantly increased.
These efforts have consequently led to the discovery of many previously-unknown VHE \gammaray sources with high 
statistical significance.  We report on the current status of this
ongoing, extended \hess GPS, present the latest images of the survey region, and highlight the
most recent discoveries.
\end{abstract}

 
\section{Introduction}
The current generation of imaging atmospheric-Cherenkov telescopes (IACTs) has
opened a new astronomical window on the Universe in the very-high-energy (VHE; E~$>$~100~GeV) domain.  Twenty
years after the detection of the first TeV \gammaray source, the Crab nebula \cite{Weekes}, approximately 100
VHE \gammaray sources have now been discovered\footnote{See TeVCat, an online TeV \gammaray catalog,
at~\url{http://tevcat.uchicago.edu}.} \cite{Hinton}.
Over two-thirds of these sources are located in our Galaxy, of which most were discovered by the IACT array \hess
(High Energy Stereoscopic System) during its ongoing Galactic Plane Survey (GPS).
VHE \gammarays carry information about the most extreme environments in the local Universe, and although a significant
fraction of the Galactic VHE \gammaray sources do not appear to have obvious counterparts at other
wavelengths \cite{AharonianDark} \cite{TibollaICRC},
the majority of them are associated with the violent, late phases of
stellar evolution, e.g. supernova remnants (SNRs), pulsar wind nebulae (PWNe) of high spin-down luminosity pulsars,
and massive Wolf-Rayet (WR) stars in stellar clusters.  Since all of these astronomical objects are known to cluster
along the Galactic plane, a comprehensive and systematic survey of this
region is an obvious approach for discovering new sources of VHE $\gamma$-rays.

\section{The \hess Telescope Array}
\label{HESSArray}

\hess is comprised of four, identical, 
12-m diameter IACTs located at an altitude of 1800 m above sea level in the Khomas Highlands of Namibia
\cite{AharonianCrab}.  Its location in the Southern Hemisphere affords it an excellent view of the inner Galaxy compared
to most of the other IACTs, which are located in the Northern Hemisphere.
Each of H.E.S.S.'s four telescopes is equipped with a camera containing 960 photomultiplier tubes 
and a tesselated mirror with a combined area of 107 m$^2$ \cite{Bernloehr}.
The optical design allows for a comparatively large, 5$^{\circ}$~field-of-view (FoV),
the largest of all the
IACTs currently in operation.
The \hess array has an angular resolution of $\sim$0.1$^{\circ}$~and an energy resolution of $\sim$15\%.  
Its unprecendented sensitivity to \gammarays above $\sim$100~GeV
enables \hess to detect a point source with a flux $\sim$1\% of the Crab nebula with a statistical
significance of 5~$\sigma$ in just 25~hours of observations \cite{AharonianCrab}.
This high sensitivity, coupled with a large FoV, permits \hess to effectively survey large areas of the Galaxy within a
reasonable amount of time.

\section{The H.E.S.S. Galactic Plane Survey}
\label{HESSGPS}

\begin{figure*}[!t] 
  \centering
  \includegraphics[width=4.73in]{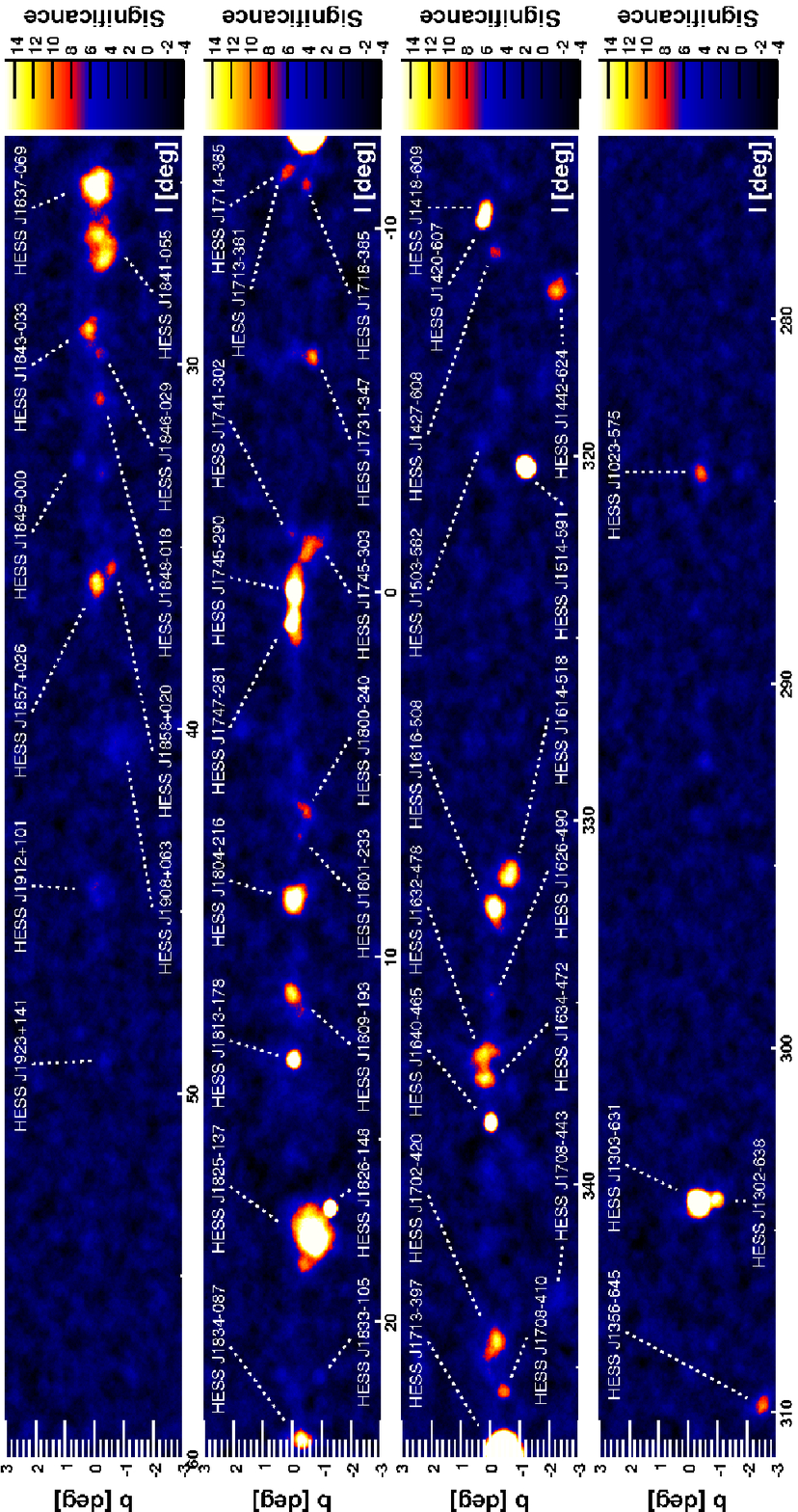}
  \caption{Image showing the pre-trials statistical significance in the \hess Galactic Plane Survey region in four
    panels.  HESS J1507$-$622 is not shown, since it is located $-$3.5$^{\circ}$~off-plane.  The
    recently-discovered VHE \gammaray emission from Westerlund~1 is also not labeled.
    The significance is truncated above 15 $\sigma$ to increase visibility, and the color transition (from blue to red)
    is set at 7.4 $\sigma$ pre-trials
    significance, which correponds to 5 $\sigma$ post-trials significance. See section \ref{ExtendedGPS}
    for more information.}
  \label{SigMap}
\end{figure*}

\subsection{The Original Survey in 2004}
The \hess Galactic Plane Survey began in 2004 and has been a major component of the \hess observation program ever since.
The original GPS \cite{AharonianGPS2}, conducted during 2004, covered the region 
$l$~$\pm$~30$^{\circ}$~in longitude and $b$~$\pm$~3$^{\circ}$~in latitude, with respect to the Galactic center (GC).
This large area (360 sq. deg.) corresponds roughly to the inner part of the Galaxy, from the Norma spiral arm tangent at 
$l$~$\approx$~327$^{\circ}$~to the Scutum-Crux spiral arm tangent at $l$~$\approx$~31$^{\circ}$~\cite{Vallee}.
Observations of 28-min duration each were taken at pointing positions separated in Galactic longitude
by $\sim$0.7$^{\circ}$.
The systematic pointings were distributed in three strips in Galactic latitude,
at $b$~$=$~$-$1$^{\circ}$, 0$^{\circ}$~and $+$1$^{\circ}$, covering
a $\sim$6$^{\circ}$-wide region along the Galactic plane.  In this \emph{scan} mode, 95 h of data were initially taken.  
Promising VHE \gammaray source candidates were then re-observed with dedicated \emph{pointed} observations,
comprising an additional 30 h
of data.  Finally, known or presumed VHE \gammaray sources were observed, including the 
GC~\cite{AharonianGC} and the shell-type SNR RX J1713.7$-$3946~\cite{Aharonian1713}.
In total, the original \hess GPS dataset included $\sim$230 h of observations after data quality selection. 

The first phase of the \hess GPS resulted in the discovery of eight, previously-unknown sources of VHE \gammarays with
a statistical significance greater than 6~$\sigma$, after accounting for all trials involved in that initial survey
(post-trials) \cite{AharonianGPS1}.  Additionally, six likely sources were detected with post-trials significances above
4~$\sigma$ \cite{AharonianGPS2}, all of which were subsequently confirmed with deeper observations.

\subsection{Extending the Survey in 2005--2009} 
\label{ExtendedGPS}
In the last four years,
the \hess GPS has been extended in Galactic longitude and now includes the 145$^{\circ}$-wide region 
$l$~$\sim$~275$^{\circ}$~--~60$^{\circ}$~\cite{ChavesGPS}.
The Survey now reaches two additional spiral arm tangents, Sagittarius-Carina at $l$~$\approx$~51$^{\circ}$~and
Crux-Scutum at
$l$~$\approx$~310$^{\circ}$~\cite{Vallee}, and beyond, for a total area of 870 sq. deg.
In addition to expanding the surveyed region, the overall exposure along the Galactic plane has also 
increased significantly.  The total exposure of the GPS dataset has increased by a factor of 6 from the initial 
$\sim$230~h of observations to over 1400~h. 
Approximately 450~h of data were taken in \emph{survey} mode, and an additional 
$\sim$950~h of data are a result of \emph{pointed} observations.

The latest image of the pre-trials statistical significance in the extended \hess GPS region is presented in
Fig.~\ref{SigMap}.
The image was created using calibrated, quality-selected data from observations during the period March 2004
to October 2008.  The on-source counts (signal plus background)
 were summed from a circle of radius 0.22$^{\circ}$~centered on each grid point.  The
background was then estimated from a ring with a mean radius of 1.2$^{\circ}$~and an area $\sim$7 times as large as the
on-source region, also centered on each grid point.
A grid spacing of 0.02$^{\circ}$ was used to produce the image.  The $\gamma$-hadron separation
was performed using \emph{hard} cuts (which require a minimum of 200 p.e. for each Cherenkov image). 
Further details on the \hess \emph{standard analysis} can be found in \cite{AharonianCrab}.  The
minimum energy threshold and the effective exposure---and, therefore, the sensitivity---vary considerably 
throughout this image.
Efforts are currently underway to increase the uniformity of the exposure across the survey region.

\begin{figure}[!t] 
  \centering
  \includegraphics[width=0.49\textwidth]{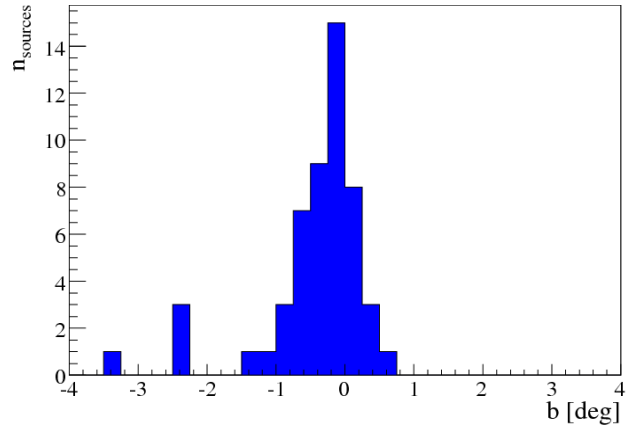}
  \caption{Distribution of the Galactic latitude $b$ of all the VHE \gammaray sources in the extended \hess GPS region.
    The distribution has a mean $b$~$=$~$-$0.26$^{\circ}$~and an rms of 0.40$^{\circ}$, not counting the four outliers at
    very low Galactic latitude.}
  \label{LatDist}
\end{figure}

The Galactic plane is clearly visible at VHE \gammaray energies, with the vast majority of VHE \gammaray sources 
distributed at low Galactic latitudes.  Figure~\ref{LatDist} shows the distribution of the \hess GPS sources in Galactic
latitude.  The distribution has a mean of $b$~$=$~$-$0.26$^{\circ}$ and an rms of 0.40$^{\circ}$, 
not counting the four outliers at $b$~$<$~2.0$^{\circ}$,
This implies the Galactic plane as seen in VHE \gammarays has a thickness of $\sim$70--120 pc in the inner Galaxy, 
compatible with the distribution of a presumed parent population of SNRs and pulsars.  However, given the non-uniform
sensitivity of the \hess GPS, this analysis is still preliminary.

The official \hess
Source Catalog is available online\footnote{See \url{http://www.mpi-hd.mpg.de/hfm/HESS/pages/home/sources}} and includes
all of the VHE \gammaray sources (Galactic and extragalactic) which were detected by \hess and subsequently published
in refereed journals.

\section{Recent Discoveries of VHE \gammaray Sources}
\label{NewSources}

\hess is continually surveying the Galaxy for VHE \gammaray source candidates and then following up on these source 
candidates with dedicated, \emph{pointed} observations.  Some of the more recent \hess discoveries are highlighted below:

\begin{itemize}
\item \textbf{HESS J1923+141}: This new source is located in the complex W~51 region, which hosts many high-energy
phenomena, among them the SNR W 51C 
(G49.2$-$0.7).  This partial shell-type SNR is interacting with a nearby giant molecular cloud (GMC),
as evidenced by the presence of two OH(1720 MHz) masers; in this scenario, VHE \gammarays could be
produced by $\pi^0$ decay after hadronic interactions in the GMC.  Competing scenarios involve leptonic interactions 
associated with a PWN candidate in the region, and, alternatively, an active star-forming region within the GMC. See
\cite{FiassonICRC} for more information.

\item \textbf{HESS J1741$-$302}: This very faint VHE \gammaray source was detected after deep observations of the GC 
region. The
plausible counterparts are currently being studied; they include cosmic ray interaction with MCs as well as a weak,
offset PWN scenario.  See \cite{TibollaICRC} for more information.

\item \textbf{HESS J1708$-$443}: One of the latest \hess discoveries in the Galactic plane
is in the vicinity of the energetic
pulsar PSR~B1706$-$44 and SNR G343.1$-$2.3.  It is an extended source of VHE \gammarays and has a relatively hard
spectrum ($\Gamma$~$=$~2.0~$\pm$~0.1$_{\mathrm{stat}}$~$\pm$~0.2$_{\mathrm{sys}}$). 
The \hess source could be associated with a relic PWN of the pulsar or with the SNR itself,
two scenarios still under investigation.  See \cite{HoppeICRC} for more information.

\item \textbf{HESS J1507$-$622}: This recently-discovered VHE \gammaray emitter is unique due to its location relatively
far from the Galactic plane, at a Galactic latitude $b$~$\sim$~$-$3.5$^{\circ}$.  It is currently unidentified
(or \emph{dark}), i.e. there are no plausible, lower-energy counterparts nearby, despite a comprehensive multi-wavelength
search.  See \cite{TibollaICRC} for more information.

\item \textbf{Westerlund 1 region}: \hess has also recently announced the discovery of VHE \gammaray emission from the
region of Westerlund 1, a massive young stellar cluster which is rich in WR stars.  See \cite{Ohm} for more information.

\end{itemize}

\section{Summary}
\label{Summary}

When the \hess Galactic Plane Survey began in 2004, there were only three known sources of VHE \gammarays in our Galaxy.
The ongoing and extended Survey now covers most of the Galactic plane as seen from the Southern Hemisphere,
specifically the region
of Galactic longitude $l$~$\sim$~275$^{\circ}$~--~60$^{\circ}$~and latitude $b$~$=$~$\pm$~3$^{\circ}$.
More than 1400 h of observations 
have now been taken in this region; these data come from a combination of systematic, $scan$-mode observations,
re-observations of promising VHE \gammaray source candidates, and dedicated observations of known or presumed VHE
\gammaray emitters.  As of May 2009, \hess has detected a total of 52 Galactic sources of VHE $\gamma$-rays in or
near the survey region.
While many of these sources are still unidentified or \emph{dark}, the majority are thought to be associated with some
of the most energetic phenomena in our Galaxy, in particular pulsar wind nebulae and supernova remnants.

\section{Acknowledgments}

The support of the Namibian authorities and of the \newpage
\noindent University
of Namibia in facilitating the construction and operation
of H.E.S.S. is gratefully acknowledged, as
is the support by the German Ministry for Education and Research (BMBF), the Max Planck Society, the French Ministry for 
Research, the CNRS-IN2P3 and the
Astroparticle Interdisciplinary Programme of the CNRS, the U.K. Science and Technology Facilities Council (STFC), the IPNP
of the Charles University, the
Polish Ministry of Science and Higher Education, the South African Department of Science and Technology and National
Research Foundation, and by the
University of Namibia. We appreciate the excellent work of the technical support staff in Berlin, Durham, Hamburg,
Heidelberg, Palaiseau, Paris, Saclay, and in
Namibia in the construction and operation of the equipment.

\end{document}